# Four fermion production in $e^+\ e^-$ annihilation [*] [†] [‡]

F. A. Berends[a] and R. Pittau[b]

Instituut-Lorentz, Leiden, The Netherlands

R. Kleiss[c]

NIKHEF-H, Amsterdam, The Netherlands

[a]e-mail: berends@rulgm0.leidenuniv.nl

[b]e-mail: pittau@rulgm0.leidenuniv.nl

[c]e-mail: t30@nikhefh.nikhef.nl

Four fermion processes at $e^+\ e^-$ colliders in a range of energy from 100 GeV to 1 TeV are discussed and examples of results obtained with an event generator presented. We also investigate the effects of the inclusion of initial state QED corrections and QCD contributions.

## 1. Introduction

Heavy vector boson production will be investigated at $e^+e^-$ colliders in a wide range of energies. At LEP II, the mass and the width of the charged boson will be measured [1] and searches for its anomalous couplings performed [2]. The relevant process is, in this case, the production of two $W$'s above threshold

$$e^+\ e^-\ \to\ W^+\ W^-\ . \qquad (1)$$

At somewhat higher energies other processes like

$$e^+e^-\ \to\ Z\ Z\ , \qquad (2)$$
$$e^+e^-\ \to\ W\ e\ \nu_e\ , \qquad (3)$$
$$e^+e^-\ \to\ Z\ e^+\ e^-\ , \qquad (4)$$
$$e^+e^-\ \to\ Z\ \nu_e\ \bar\nu_e\ , \qquad (5)$$

become important [3].

Since the produced heavy bosons immediately decay, for all above processes the detected experimental signal is a four fermion final state, which has important consequences. First of all, the number of possible final states grows. Secondly, the number of Feynman diagrams contributing to a given final state can be very large. Finally, the same four fermion final state can originate from several signals. For example, all 5 signals (1)-(5) can end up with a $e^+e^-\nu_e\bar\nu_e$ final state.

Of course, at some energies, a particular set of diagrams dominates, and all the others can be considered as a background. At LEP II, three diagrams with 2 decaying $W$'s dominate, while, at next linear colliders in the TeV regime, diagrams for single W or Z production, are responsible for the leading effects and the former diagrams give small contributions. For an exhaustive treatment of all questions concerning signal and background see ref. [4]. Here, we only stress that it is not always possible to use kinematical cuts to remove the background effects. For instance, if two neutrinos are present in the final state, no invariant mass cuts can be performed and only quantities summed over all possible neutrino flavours are suitable for comparisons with experiment.

Thus, an explicit calculation of *all* four fermion processes is unavoidable, in order to check the effectiveness of cuts to reduce the background and to extract physical information whenever these cuts are not applicable.

This rather ambitious program has been carried out in five papers [4]-[8]. The main result is a very fast Monte Carlo program, **EXCALIBUR** [5,6], that, in the limit of vanishing fermion masses, takes into account all possible Feynman diagrams contributing to any given four fermion final state.

---

[*]Talk presented by R. Pittau at Zeuthen meeting in April 1994.

[†]Research supported in part by the EU under contract n. CHRX-CT-0004.

[‡]Preprint INLO-PUB-7/94.

Note that neglecting fermion masses implies the *absence* of diagrams where a Higgs boson couples to the fermions: but, although we cannot compute the Higgs signal in our approach, we can at least reliably estimate the background. For one diagram in the background one then has to introduce a special treatment, i.e. the diagram where two $W'$s fuse to $b\bar{b}$ with top quark exchange, which has to be taken massive.

The QCD tree level effects in 4 jets final states [7] and the initial state QED radiation [8] are also implemented, so we feel confident that our program is complete and accurate enough to describe, in the framework of the Standard Model, the four fermion physics relevant at $e^+e^-$ colliders.

The outline of this paper is as follows. In section 2 we sketch the method used in computing the matrix element and in building the Monte Carlo. In section 3, 4 and 5 results obtained with the described generator are presented. Finally, we summarize our conclusions.

## 2. The method

Since one wants to deal with all possible final states, first of all an efficient way to compute the matrix elements is needed.
There are Abelian and non Abelian graphs with two distinct topologies [4]. All Feynman diagrams can be obtained from those two topologies by permuting the fermion momenta. The number of diagrams can be very large (up to 144), so the evaluation has to be performed at the level of helicity amplitudes. Using spinorial techniques, one single basic function can be found that describes the spinorial structure of both Abelian and non Abelian graphs. Thus, for massless fermions, each helicity amplitude consists of a sum of very systematic and relatively compact expressions. This is important for the computational speed.

As for the Monte Carlo integration over the final fermion momenta, we have used a *multi-channel* approach. The basic idea is to generate the integration variables according to distributions that approximately reproduce the peaking behaviour of the matrix element squared. This reduces the variance of the integrand, and therefore the Monte Carlo error.

Since the peaking structure of the matrix element squared can be very rich, one set of integration variables can only describe a limited number of peaks, so that, in general, one has to generate each peak with different mappings of random numbers, that is different *channels*. In EXCALIBUR those *channels* are automatically chosen by the program, looking at the Feynman diagrams contributing to the selected process. Also the optimization of the predetermined probabilities used to choose the various *channels* is automatic, leading to an effective increase in the program speed by almost an order of magnitude [6].

QED corrections are implemented using the structure-function formalism [9]. Each of the incoming fermions is assumed to have its energy degraded by the emission of photons parallel to the beam. For the energy distribution of the fermion after radiation we take a structure function $\Phi$ that incorporates the leading log $\mathcal{O}(\alpha)$ and $\mathcal{O}(\alpha^2)$ initial state radiation with exponentiation of the remaining soft-photon effects [8]. Our model for the total radiative cross section is then

$$\sigma(s) = \int_0^1 \int_0^1 dx_1\, dx_2\, \Phi(x_1)\, \Phi(x_2)\, \sigma_0(x_1 x_2 s) \quad (6)$$

where $\sigma_0$ is the nonradiative cross section and $x_1$, $x_2$ represent the energy content of the incoming fermions after radiative emission. This provides an adequate description of the leading QED effects.

When four quarks are present in the final state, one has to add the concomitant QCD production channels, and also the production of a quark pair and two gluons, since both types of final states will appear as jets. In our Monte Carlo the former contribution is easily implemented without additional CPU time by adding gluons wherever photons connect quark lines [7] (of course the correct QCD coupling and colour structure should be taken into account). Finally, the latter process can be efficiently computed using the recursion relations of ref. [10].

Table 1
Cross section in pb for $e^+e^- \to e^+e^- \nu_{e,\mu,\tau} \bar\nu_{e,\mu,\tau}$. Cuts applied: $m_{(e^+e^-)} > 10\ GeV$, $|\cos\theta_{e^\pm}| < 0.9$, $E_{e^\pm} > 20\ GeV$.

| $\sqrt{s}$ (GeV) | 150 | 175 | 200 |
|---|---|---|---|
| WW diagrams only | $.3600 \pm .0011\ 10^{-2}$ | $.1181 \pm .0002$ | $.1304 \pm .0003$ |
| WW + background | $.4625 \pm .0011\ 10^{-2}$ | $.1246 \pm .0002$ | $.1591 \pm .0003$ |
| Relative difference (%) | 22 | 5 | 18 |

Table 2
Cross section in pb for $e^+e^- \to e^- \bar\nu_e + 2\ jets$. Cuts applied: $E_{e^-,\ q,\ \bar q'} > 20\ GeV$, $|\cos\theta_{e^-,\ q,\ \bar q'}| < 0.9$, $|\cos_{(q,\bar q')}| < 0.9$, $m_{(q\bar q')} > 10\ GeV$.

| $\sqrt{s}$ (GeV) | 150 | 175 | 200 |
|---|---|---|---|
| WW diagrams only | $.1921 \pm .0060\ 10^{-1}$ | $.6684 \pm .0014$ | $.7202 \pm .0016$ |
| WW + background | $.1513 \pm .0042\ 10^{-1}$ | $.6690 \pm .0014$ | $.7490 \pm .0018$ |
| Relative difference (%) | -27 | 0.1 | 4 |

## 3. Results without initial state radiation

In this section we shall present examples of numbers obtained with **EXCALIBUR**. We start showing results without QED radiation and QCD effects [4]. The actual values for the input parameters are $\alpha = 1/129$, $\sin^2\theta_W = 0.23$, $M_W = 80.5$, $\Gamma_W = 2.3$, $M_Z = 91.19$ and $\Gamma_Z = 2.5$ (all GeV).

In the first row of table 1 only the three WW signal diagrams are considered, while in the second entry all diagrams leading to a final state with $e^+e^-$ and two neutrinos are taken into account. All neutrino flavours are summed over, because, as already discussed in the introduction, it is not possible to distinguish them experimentally. As a consequence, the background for this process consists of both an *interfering* and a *non-interfering* part. The last entry shows the whole effect. Usually, one tries to get rid of the background contributions by imposing invariant mass cuts on the particles that can come from decaying $W$'s. Note that this is not possible here because two neutrinos are produced. Therefore, any study on, say, lepton universality made at LEP II on the basis of purely leptonic events like those shown in table 1, should take care of the background. A last remark is in order. While a large background below threshold (22%) is somewhat understandable, the large contribution at 200 GeV (18%) may appear surprising. This is due to two combined effects. On one hand diagrams for single $W$ production become important at higher energies. On the other hand, at 200 GeV, background $Z$'s are produced above threshold.

In table 2 the semileptonic process $e^+e^- \to e^- \bar\nu_e + 2\ jets$ is considered (taking the CKM matrix unity it is the sum of cross sections with $u\bar d$ and $c\bar s$ in the final state). Here, only *interfering* background is possible. Then, since no resonant $Z$'s are produced, at 200 GeV, the effect of the background is not as large as in the previous case. For the process $e^+e^- \to \mu^- \bar\nu_\mu + 2\ jets$, where also diagrams with single $W$ production are absent, the contribution of the background is even smaller (0.3%), again at 200 GeV. With semileptonic final states invariant mass cuts are possible. A cut $70 < m_{(e^-\bar\nu_e)},\ m_{(q\bar q')} < 90$ GeV reduces the background from 4 % to 1% for the process in table 2 at 200 GeV, and only 10% of the events are lost.

Finally, in table 3, the influence of the *interfering* electroweak background on a specific four quark final state is investigated.

This concludes our discussion on signal and background at LEP II. At higher energies single boson production processes become important and the problem is to find a way to disentangle, for any given final state, the signals (1)-(5). In ref. [4] we showed that this can be achieved by cutting the invariant masses of the events around the vector boson masses. For example, at 1 TeV and with the cuts $E_{e^-,\ u,\ \bar d} > 20\ GeV$, $|\cos\theta_{e^-,\ u,\ \bar{d}}| < 0.9$, $m_{(u\bar{d})} > 30\ GeV$, the cross section for the process $e^+e^- \to e^-\bar{\nu}_e u\bar{d}$ including *all* diagrams is 0.0452 pb, while the result with *only WW diagrams* is 0.0143 pb. That is a difference of a factor 3. An additional cut $M_W - 2\Gamma_W < m_{(e^-\bar{\nu}_e)}$, $m_{(u\bar{d})} < M_W + 2\Gamma_W$ reduces *both* numbers to 0.0079 pb. However, even if such a procedure removes all the other signals, it should be noted that almost half of the $WW$ signal has been cut. Therefore, in order to avoid this huge loss in statistics, any study about tiny effects (for instance searches for anomalous couplings) has to be performed directly at the level of a *complete* $e^+e^- \to 4$ fermions generator.

4Table 3
Cross section in pb for $e^+e^- \to u\bar{d}d\bar{u}$. $E_{(all\ particles)} > 20\ GeV$, $|\cos\theta_{(all\ particles)}| < 0.9$. Moreover $m_{(ij)} > 10\ GeV$ and $|\cos(i,j)| < 0.9$ between all possible final state pairs.

| $\sqrt{s}$ (GeV) | 150 | 175 | 200 |
|---|---|---|---|
| WW diagrams only | $.2141 \pm .0019\ 10^{-1}$ | $.7699 \pm .0018$ | $.8726 \pm .0023$ |
| WW + background | $.2317 \pm .0010\ 10^{-1}$ | $.7745 \pm .0018$ | $.8987 \pm .0023$ |
| Relative difference (%) | 8 | 0.6 | 3 |

Table 4
Results on radiatively corrected cross sections and average energy losses, under various calculational strategies, for the process $e^+e^- \to e^-\bar{\nu}_e u\bar{d}$. The cross sections are given in picobarns, $\bar{\epsilon}$ and $\epsilon$ in GeV.

| | $\sqrt{s}=176$ GeV | | | |
|---|---|---|---|---|
| strategy | $\sigma_0$ | $\sigma$ | $\bar{\epsilon}$ | $\epsilon$ |
| **WW,f,1** | $.60111 \pm .00032$ | $.50490 \pm .00032$ | $1.162 \pm 0.002$ | – |
| **WW,s,1,a** | " | $.50484 \pm .00033$ | $1.172 \pm 0.002$ | $1.175 \pm 0.002$ |
| **WW,s,1,b** | " | $.50175 \pm 00098$ | $1.167 \pm 0.006$ | $1.170 \pm 0.006$ |
| **WW,s,2** | " | $.50258 \pm .00097$ | $1.178 \pm 0.006$ | $1.181 \pm 0.006$ |
| **WW,cuts** | $.44651 \pm .00092$ | $.37737 \pm .00091$ | $1.192 \pm 0.007$ | $1.195 \pm 0.007$ |
| **all,cuts** | $.45011 \pm .00097$ | $.37926 \pm .00095$ | $1.149 \pm 0.007$ | $1.152 \pm 0.007$ |

## 4. Results with initial state radiation

In order to present some salient results we have chosen the specific process

$$e^+e^- \to e^-\bar{\nu}_e u\bar{d}\ , \qquad (7)$$

which contains the $WW$ pair production signal, as well as a non-negligible background [8].

Table 4 contains numbers from our Monte Carlo obtained under different strategies. The quoted values are the nonradiative cross section $\sigma_0$, the total cross section with initial state radiation $\sigma$, the average 'energy loss' $\bar{\epsilon}$ defined as $\frac{\sqrt{s}}{2}(1 - x_1 x_2)$ (see ref. [11]) and the *real* energy loss $\frac{\sqrt{s}}{2}(2 - x_1 - x_2)$, denoted by $\epsilon$. The strategies used are

- **WW,f,1**: the $WW$ signal diagrams only, with the flux-function approach to $\mathcal{O}(\alpha)$ as given in ref. [11]; no cuts;

- **WW,s,1,a**: the $WW$ signal diagrams only, with structure functions that are simply the flux function of ref. [11] to $\mathcal{O}(\alpha)$, in which $2\alpha$ is replaced by $\alpha$; no cuts;

- **WW,s,1,b**: the $WW$ signal diagrams only, with $\mathcal{O}(\alpha)$ structure functions of ref. [8]; no cuts;

- **WW,s,2**: the $WW$ signal diagrams only, with $\mathcal{O}(\alpha^2)$ structure functions as given in ref. [8]; no cuts;

- **WW,cuts**: like the previous case, except that now we also impose the following cuts: $E_{e^-,\ u,\ \bar{d}} > 20\ GeV$, $|\cos\theta_{e^-,\ u,\ \bar{d}}| < 0.9$, $|\cos(u\bar{d})| < 0.9$, $m_{(u\bar{d})} > 10\ GeV$;

- **all,cuts**: like the previous case, except that now also all the background Feynman diagrams are taken into account.

Table 5
Cross section in pb for $e^+e^- \to 4\ jets$. Cuts as in table 3.

| $\sqrt{s}$ (GeV) | 175 | 190 |
|---|---|---|
| WW signal | $3.0674 \pm 0.0074$ | $3.5136 \pm 0.0090$ |
| WW signal + ISR | $2.5622 \pm 0.0071$ | $3.1416 \pm 0.0089$ |
| All EW diagrams + ISR | $2.5922 \pm 0.0075$ | $3.3553 \pm 0.0097$ |
| All EW diagrams + interfering QCD + ISR | $2.6202 \pm 0.0079$ | $3.3789 \pm 0.0100$ |
| All EW diagrams + all QCD + ISR | $3.1155 \pm 0.0123$ | $3.8688 \pm 0.0146$ |

For this analysis we have adopted the values given in ref. [11], namely

$$\alpha = (127.29)^{-1}\ ,\quad \sin^2\theta_w = 0.2325\ ,$$
$$\alpha_{\text{str.f., flux}} = (137.036)^{-1}\ ,$$
$$M_Z = 91.173\,\text{GeV}\ ,\quad \Gamma_Z = 2.4971\,\text{GeV}\ ,$$
$$M_W = 80.220\,\text{GeV}\ ,\quad \Gamma_W = 2.033\,\text{GeV}\ .$$

Referring to eq. 6, $\alpha$ is contained in $\sigma_0$, while $\alpha_{\text{str.f., flux}}$ is used in $\Phi(x)$. The structure function method differs slightly from the flux function method as comparisons between the first and second row of table 4 show. The inclusion of cuts or more diagrams affect both cross sections and energy losses.

Since the proposed direct reconstruction method for the $W$ mass suffers from a shift in $M_W$ due to the radiated energy [1], a precise knowledge of $\bar{\epsilon}$ and $\epsilon$ is warranted. Our results show that the precise treatment of ISR, the choice of cuts and the neglect of diagrams all affect the energy losses. The first effect was also found in ref. [11], where also the influence of the Coulomb singularity is discussed. The two other effects show that a Monte Carlo treatment allowing for cuts and being able to include all diagrams is indispensable.

## 5. Results with initial state radiation and QCD contributions

With four-quark final states also the QCD contributions must be taken into account [7]. Table 5 shows results for the process $e^+e^- \to 4\ jets$ (neglecting fragmentation effects) under inclusion of several QED and QCD contributions at LEP II energies. The input parameters are as in section 3 and $\alpha_s = 0.103$. The cross section is lowered by the initial state QED radiation (ISR), while all the other contributions tend to raise it back to its Born value. The electroweak (EW) background is at the per cent level as well as the interfering QCD background (four quark production via gluons). In the last row also the non-interfering QCD cross section $e^+e^- \to 2$ gluons + 2 quarks is included; it increases the cross section by 16% at $\sqrt{s}=$ 175 GeV.

In figure 1 the distribution of the invariant masses for the WW signal, the fully corrected cross section and the non-interfering QCD contribution are shown at $\sqrt{s}=$ 175 GeV. The curves have been obtained by fitting histograms from EXCALIBUR. The combinatory background is included too, by taking into account any possible invariant mass $m_{(i,j)}$, so that the area under each curve is 6 times the value of the corresponding cross section. ISR lowers the distributions and $ggq\bar{q}$ gives a rougly constant positive contribution between 30 and 100 GeV. Below the peak the latter effect is dominant, while at the peak the situation is reversed. A more detailed analysis can be found elsewhere [7]. Here, we only report that the maximum and the width are not affected by QED initial state radiation. On the other hand, the QCD and combinatory background increase the width.

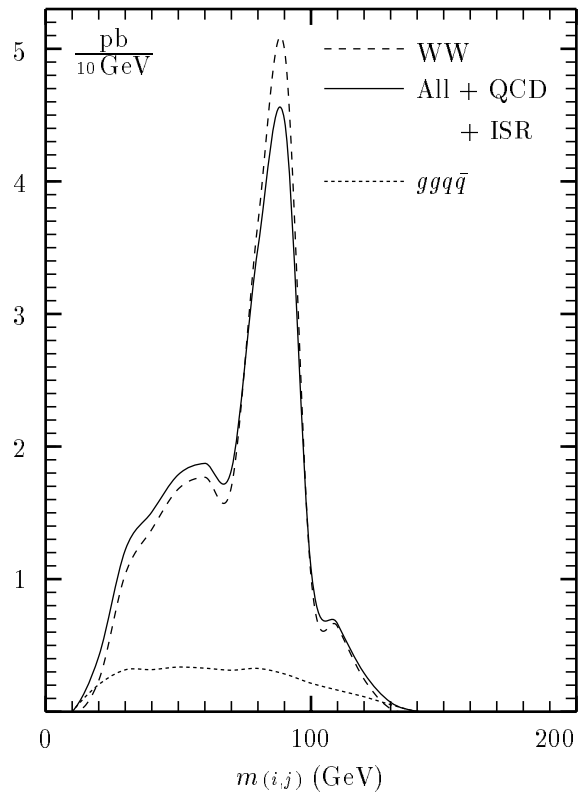

Figure 1
Distribution of the invariant masses for $e^+e^- \to 4\ jets$ at $\sqrt{s} = 175$ GeV.

## 6. Conclusions

Four fermion production in $e^+e^-$ annihilation will allow important tests of the Standard Model and measures at LEP II and beyond.

There are processes and experimental setups for which taking into account only subclasses of Feynman diagrams is a good approximation, but this is not always the case. As a consequence, programs including *all* diagrams and being able to compute *any* possible four fermion final state are indispensable. Also initial state QED radiation and QCD contributions turn out to be important, and must be taken into account. EXCALIBUR is a Monte Carlo program satisfying all these requirements. A Monte Carlo approach is unavoidable, since important effects (e.g. the average energy loss) are very sensitive to the imposed experimental cuts.